# Transition Frequencies and Dynamic Amplification of Buried Lifelines: A Semi-Analytical Timoshenko Beam on Winkler Foundation Model

Gersena Banushi[1] and Kenichi Soga[1]

[1] Department of Civil and Environmental Engineering, University of California, Berkeley, Berkeley, CA 94720, USA
g.banushi@berkeley.edu

**Abstract.** Underground lifelines, such as pipelines and tunnels, are susceptible to ground vibrations from seismic events, traffic, and other dynamic sources. Accurate prediction of their response is essential for ensuring structural safety and operability. This study introduces a semi-analytical model for transverse vibration analysis of buried lifelines, formulated using the Timoshenko beam theory on elastic foundation. The closed-form analytical solutions revealed that the vibration spectrum comprises four parts, separated by three transition frequencies. At each transition, the oscillatory characteristics of the modes change as a function of the system properties, leading to marked variations in dynamic amplification. The model's validity is confirmed through case studies of buried steel pipelines of varying lengths and operating conditions, showing excellent agreement with finite element simulations. A subsequent parametric study quantifies the influence of key factors – including soil stiffness and system length – on dynamic performance. The proposed method provides a computationally efficient and physically transparent framework for capturing complex vibration phenomena beyond simplified travelling-wave approaches, offering valuable guidance for the design and resilience assessment of underground lifeline systems subjected to various dynamic loads.

.

# 1    Introduction

Underground lifelines, such as pipelines and tunnels, constitute the arteries of modern infrastructure systems, ensuring continuous supply of energy, water, and essential services. Their structural integrity and operability are critical to ensure system safety and resilience particularly under dynamic excitations caused by seismic events, traffic-induced vibrations, and environmental loading.

Current seismic design methods for buried pipelines [1-3] are based on simplistic analytical models that idealize the soil movement as a traveling sinusoidal wave, while ignoring the system's inertia and the relative soil-structure movement. However, this assumption may not be valid for buried large diameter pipelines and tunnels requiring accurate dynamic analysis of the buried structure subjected to transient ground deformation (TGD) in the axial and transverse directions.

Advanced analytical and numerical approaches account for soil–structure interaction, with the beam-on-Winkler-foundation model being one of the most widely applied in lifeline engineering, including railway tracks, tunnels, and buried pipelines. Herein, the system response has been analyzed either quasi-statically, i.e. by ignoring the inertia effects, or dynamically [4].

The dynamic analysis of buried beams includes the free structural vibration and the forced transient response, using either the Euler-Bernoulli or the Timoshenko beam theory. The former approach has been more widely employed in dynamic analysis of lifelines resting on Winkler foundation, because of the simpler mathematical formulation [4-6]. On the other hand, the Timoshenko model considers the effect of transverse shear deformation and rotary inertia on the beam modes and frequencies, providing a more accurate dynamic response. Among various boundary conditions, the free-free Timoshenko beam on elastic foundation has been less investigated, either analytically or numerically, because of the more complex computation, including the rigid-body motion [7].

To analyze the dynamic response of a buried straight beams with free ends subjected to transverse TGD, this study adopts a new semi-analytical model introduced in [8], based on the Timoshenko beam on Winkler foundation theory. The model can evaluate the free vibration and forced dynamic response for various system lengths and operating conditions. The application of the model is illustrated by simulating the case of a buried straight pipeline with variable lengths and soil conditions. The correctness of the analytical solution is verified using finite element modal analysis and implicit dynamic simulations. The implemented model is further used to investigate the influence of the main system parameters impacting dynamic behavior for a better understanding of the system response.

# 2    Methodology

This section describes the methodology for evaluating the dynamic response of a Timoshenko beam on Winkler foundation, through modal analysis. First, the differential equation of vibration of a Timoshenko beam on Winkler foundation is solved in a closed form solution, using the Fourier method of variable separation. Second, the natural vibration frequencies and corresponding modal shapes are evaluated for the whole spectrum, considering free-free boundary conditions. Third, the vibration spectrum is constructed explicitly using the case study of a Timoshenko beam on Winkler foundation, modeling a buried pipeline subjected to seismic wave propagation. The calculated analytical solutions in terms of vibration frequencies and modes for various beam lengths are compared with the finite element modal analysis results. The forced dynamic response of the Timoshenko beam on Winkler foundation is evaluated by solving the modal equations in the unknown modal coordinates numerically, and then calculating the total displacement as a linear combination of all modes. The obtained response is compared with the finite element analysis results, confirming the validity of the analytical method. The implemented model is further used to perform a frequency response analysis for various beam lengths and soil conditions, providing a deeper understanding of the system dynamic behaviour.

To investigate the dynamic response of a Timoshenko beam on elastic foundation, we consider the case study of a straight steel gas pipeline with variable length buried in medium dense sand, as shown in Figure 1a. The investigated steel pipeline has a diameter $D = 107$ cm (42 in.) and wall thickness $t = 2.2$ cm (7/8 in.) and is assumed buried in medium dense sand at a depth $H = 1.45$ m. Table 1 summarizes the soil-structure system parameters used in this study.

The numerical analysis of the buried pipeline subjected to ground vibration was performed using the finite element software ABAQUS/Standard [9]. The buried pipeline was modeled using PIPE21H beam elements, while the soil-structure interaction was represented with the spring-like pipe-soil interaction elements PSI24. One edge of the soil element shares nodes with the underlying pipe element while the nodes on the other edge are assigned the base ground motion with a time shift proportional to the distance along the system axis (Figure 1a). A 1.0 m mesh size was adopted for both pipe and soil elements, based on a mesh sensitivity study ensuring solution accuracy and computational efficiency.

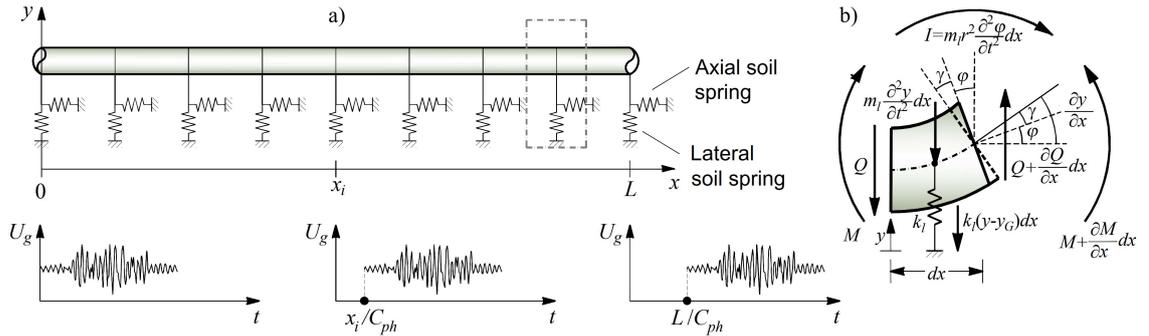

**Fig. 1.** Schematic representation of: a) the buried Timoshenko beam model subjected to seismic-induced ground displacement time histories with a time shift at axial distance $x_i$ equal to $x_i/C_{ph}$; and b) the translational and rotational equilibrium of a differential beam element of length $dx$ under lateral deformation, leading to the governing vibration equation of the Timoshenko beam on elastic foundation [8].

**Table 1.** Soil-structure system parameters.

| Parameter | Value |
| --- | --- |
| Pipeline diameter, $D$ (m) | 1.067 |
| Pipeline wall thickness, $t$ (m) | 0.022 |
| Pipeline burial depth, $H$ (m) | 1.45 |
| Soil density, $\gamma$ (kN/m$^3$) | 17.0 |
| Soil friction angle, $\phi$ (°) | 30 |
| Soil friction reaction per unit pipe length, $f_r$ (kN/m) | 41.22 |
| Axial soil-spring maximum elastic deformation, $u_0$ (cm) | 0.15 |
| Linear axial soil-spring stiffness, $k_a = 2f_r/u_0$ (MPa) | 56.044 |
| Lateral soil reaction per unit pipe length, $p_u$ (kN/m) | 123.98 |
| Lateral soil-spring maximum elastic deformation, $u_l$ (cm) | 9.91 |
| Lateral soil-spring stiffness, $k_l = 2p_u/u$ (MPa) | 2.503 |
| Steel elastic modulus, $E$ (GPa) | 210 |
| Steel Poisson's ratio, $\nu$ | 0.3 |
| Elastic shear modulus, $G = 0.5E/(1 + \nu)$, (GPa) | 80.77 |
| Beam shear coefficient, $k$ | 0.53 |
| Beam cross-sectional area, $A_b$ (m$^2$) | 0.026 |
| Beam second moment of inertia, $J$ (m$^4$) | 0.0037 |
| Radius of gyration of the beam cross section, $r = \sqrt{J/A_b}$ (m) | 0.374 |
| Steel density, $\rho$ (kg/m$^3$) | 7860.35 |
| Linear mass of the steel pipeline, $m_l$ (kg/m) | 573.29 |

The numerical dynamic modal analysis is conducted in two consecutive steps. First, the eigenvalue extraction of the first 1000 modes is performed, calculating the corresponding natural frequencies and the mode shapes of the system. Second, in the modal dynamic step, the modal amplitudes are integrated through time, and the dynamic response is obtained by modal superposition.

## 3    Semi-Analytical Timoshenko Beam on Winkler Foundation Model

This section describes the semi-analytical model introduced in [8] for evaluating the dynamic response of a buried straight beam subjected to transverse TGD. The equation of vibration of the Timoshenko beam resting on elastic foundation is obtained considering transitional and rotational equilibrium of a differential beam element subjected to lateral excitation, as schematically illustrated in Figure 1b:

$$EJ\frac{\partial^4 y}{\partial x^4} - m_l r^2\left(1+\frac{E}{kG}\right)\frac{\partial^4 y}{\partial x^2 \partial t^2} + m_l\frac{\partial^2 y}{\partial t^2} + \frac{m_l^2 r^2}{kGA_b}\frac{\partial^4 y}{\partial t^4} +$$
$$+ k_l\left(y - \frac{EJ}{kGA_b}\frac{\partial^2 y}{\partial x^2} + \frac{m_l r^2}{kGA_b}\frac{\partial^2 y}{\partial t^2}\right) = k_l\left(y_G - \frac{EJ}{kGA_b}\frac{\partial^2 y_G}{\partial x^2} + \frac{m_l r^2}{kGA_b}\frac{\partial^2 y_G}{\partial t^2}\right) \quad (1)$$

where $m_l$ is the linear mass of the beam per unit length, $EJ$ is the beam flexural rigidity, $G$ is the modulus of rigidity, $A_b$ is the cross sectional area, $r = \sqrt{J/A_b}$ is the radius of gyration of the beam cross section and $k$ is a constant that depends on the cross-sectional shape and accounts for the nonuniform distribution of shear stress across the section, and $k_l$ is the rigidity of the lateral soil-structure interaction, $y_G$ is the transient ground displacement, as shown in Figure 1b.

The Fourier method of variable separation is used to calculate the beam displacement response $y(x,t)$, as a linear combination of the displacements of all vibration modes $y_n(x,t)$:

$$y(x,t) = \sum_{n=1}^{\infty} y_n(x,t) = \sum_{n=1}^{\infty} \phi_n(x) q_n(t) \quad (2)$$

where, $q_n(t)$ and $\phi_n(x)$ are the modal coordinate and modal shape corresponding to the $n$-th natural vibration frequency $\omega_n$, respectively.

Substituting Eq.(2) into Eq.(1), allows to obtain the fourth order differential equation governing the spatial functions $\phi_n(x)$. Its solutions have the form of exponential functions, $\exp(\lambda x)$, where, in general, $\lambda \in \mathbb{C}$ [8]. The nature of the roots $\lambda$ varies across the three transition frequencies $\tilde{\omega}_i$ in Eq. (3), significantly influencing the oscillatory characteristic of the solution $\phi_n(x)$, as summarized in Table 2.

$$\tilde{\omega}_1^2 \cong \frac{k_l}{m_l}\left(1 - \frac{1}{2r^{-4}(EJ/k_l) + (1 - E/kG)}\right), \quad \tilde{\omega}_2^2 = \frac{k_l}{m_l}, \quad \tilde{\omega}_3^2 = \frac{kGA_b}{m_l r^2} \quad (3)$$

In Eq. (3), the term $EJ/k_l$ denotes the flexibility of the buried beam relative to the soil stiffness, which becomes large for beams resting on soft foundations, and results in the first transition frequency $\tilde{\omega}_1$ *being* very close to the second, $\tilde{\omega}_2$.

**Table 2.** Modal shape $\phi_n(x)$ for the different parts of the frequency vibration spectrum.

| Frequency | Modal shape $\phi_n(x)$ | Eigenvalues $\lambda_{1/2}$ |
|---|---|---|
| $\omega_n < \tilde{\omega}_1$ | $\phi_n(x) = e^{-\alpha x}(C_1\sin\beta x + C_2\cos\beta x) + e^{\alpha x}(C_3\sin\beta x + C_4\cos\beta x)$ | $\lambda_{1/2} = \alpha \pm i\beta$ |
| $\omega_n = \tilde{\omega}_1$ | $\phi_n(x) = C_1\sin\beta x + C_2\cos\beta x + x(C_3\sin\beta x + C_4\cos\beta x)$ | $\lambda_1 = \lambda_2 = \pm i\beta$ |
| $\tilde{\omega}_1 < \omega_n < \tilde{\omega}_2 \vee \omega_n > \tilde{\omega}_3$ | $\phi_n(x) = C_1\sin\beta_1 x + C_2\cos\beta_1 x + C_3\sin\beta_2 x + C_4\cos\beta_2 x$ | $\lambda_1 = \pm i\beta_1, \lambda_2 = \pm i\beta_2$ |
| $\omega_n = \tilde{\omega}_2 \vee \omega_n = \tilde{\omega}_3$ | $\phi_n(x) = C_1 + C_2 x + C_3\sin\beta x + C_4\cos\beta x$ | $\lambda_1 = 0, \lambda_2 = \pm i\beta$ |
| $\tilde{\omega}_2 < \omega_n < \tilde{\omega}_3$ | $\phi_n(x) = C_1\sinh\alpha x + C_2\cosh\alpha x + C_3\sin\beta x + C_4\cos\beta x$ | $\lambda_1 = \pm\alpha, \lambda_2 = \pm i\beta$ |

These solutions contain four unknown constants $C_1$, $C_2$, $C_3$, $C_4$, and the eigenvalue parameters $\alpha$ and $\beta$. Application of the four end boundary conditions for a single-span beam provides a solution for the natural frequency $\omega_n$, and for the three constants in terms of the fourth, resulting in the natural mode shapes $\phi_n(x)$.

## 4 Dynamic Response of buried Timoshenko Beams

### 4.1 Free vibration

Figure 3 illustrates the full frequency spectrum for the first 100 modes, highlighting the number of modes for each part of the spectrum, considering the steel pipeline buried in compacted and poorly compacted backfill. The former presents higher vibration frequencies than the latter, because of the greater soil stiffness. Clearly, the number of vibration modes in the low frequency range $N(\omega_n \leq \tilde{\omega}_2)$ as well as the modal density increase with the system length $L$ and soil stiffness $k_l$.

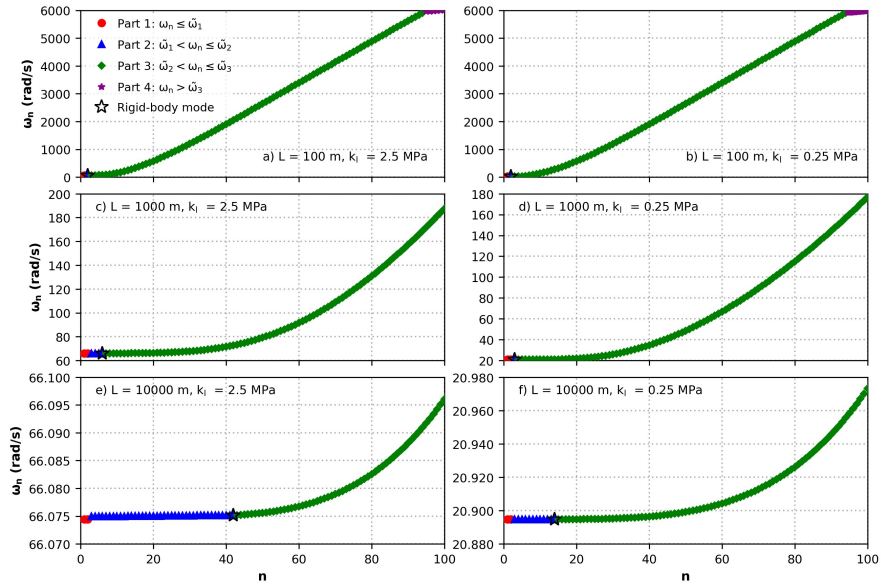

**Fig. 3.** Frequency spectrum representing the first 100 vibration frequencies $\omega_n$ versus the mode number $n$ for different beam lengths and soil conditions: $L = 100$ m, a) compacted ($k_l = 2.5$ MPa) and b) poorly compacted soil ($k_l = 0.25$ MPa); $L = 1000$ m: c) compacted and d) poorly compacted soil; $L = 10000$ m: e) compacted and f) poorly compacted soil.

The evaluated modal shapes for the first, second, and third part of the frequency spectrum, are shown in Figure 4, considering the proposed analytical and finite element modal analysis. Comparison between the semi-analytical and numerical results shows excellent agreement between the two approaches, further confirming the validity of the implemented semi-analytical model.

### 4.2 Forced vibration under transient ground deformation (TGD)

The ground motion $U_g(x,t)$ is modelled as a sinusoidal shear wave with amplitude $U_{g,max}$, propagating with an angular frequency $\omega$ and apparent wave velocity $C_{ph} = 2000$m/s [8]. In the semi-analytical approach, 10, 90, and 220 modes were required for the 100 m, 1000 m, and 10000 m long system, respectively.

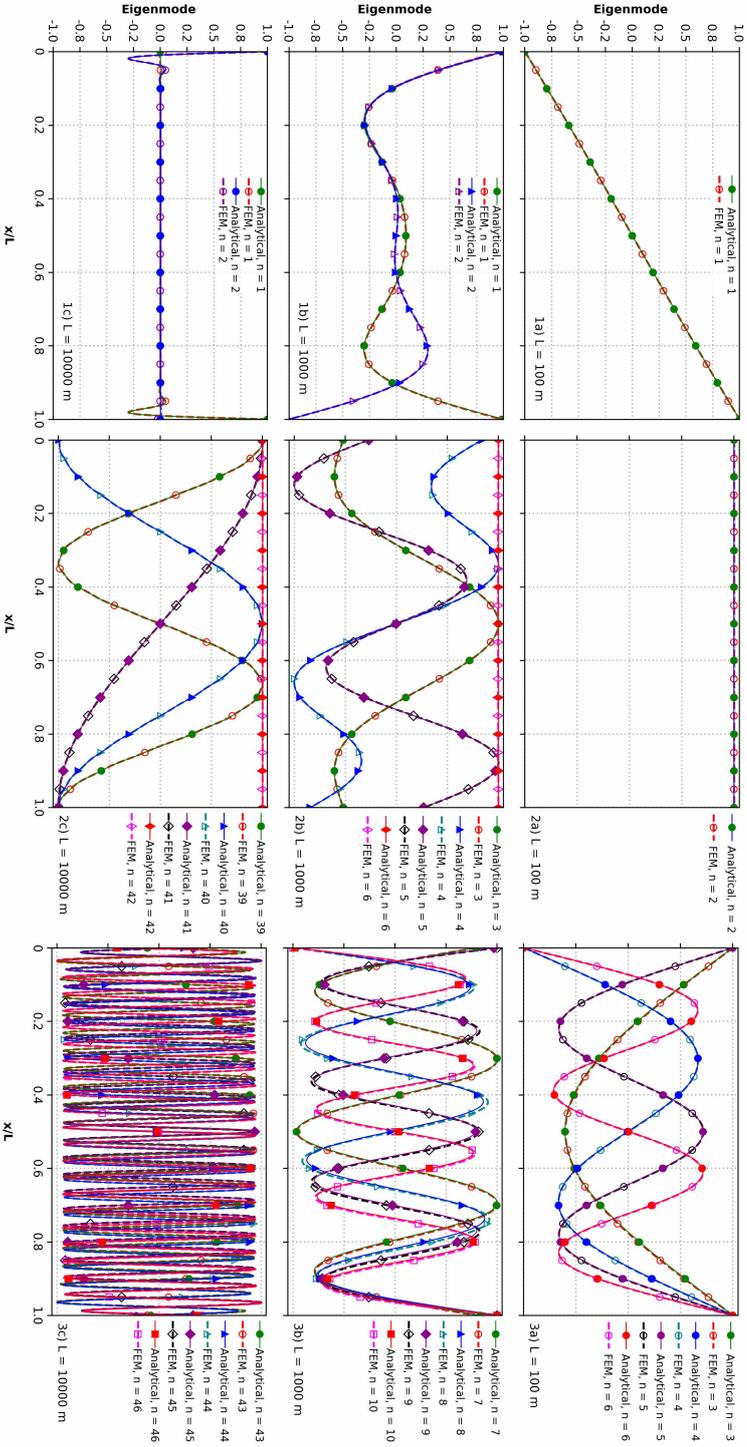

**Fig. 4.** Natural vibration modes for the different parts of the frequency spectrum $i$ ($i = 1$: $\omega_n \leq \tilde{\omega}_1$; $i = 2$: $\tilde{\omega}_1 < \omega_n \leq \tilde{\omega}_2$; $i = 3$: $\tilde{\omega}_2 < \omega_n \leq \tilde{\omega}_3$) and various system lengths $L$: $ia$) $L = 100$ m; $ib$) $L = 1000$ m; $ic$) $L = 10000$ m.

Figure 5 shows the displacement of the pipeline buried in compacted soil at $t = 5$ s, obtained using the semi-analytical, finite element modal and implicit dynamic analysis, assuming a ground motion frequency of 0.5 Hz. Excellent agreement between the analysis approaches is apparent. The total pipeline vibration is the sum of two harmonic curves, characterized by the ground motion forcing frequency (0.5 Hz), and the fundamental frequency of the buried pipeline (10.5 Hz). The former is smaller than the latter, leading to a quasi-static system response.

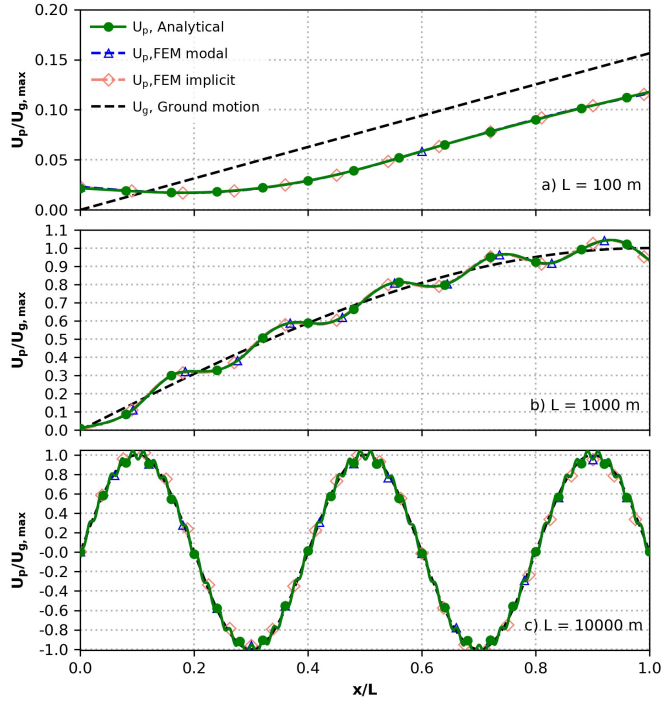

**Fig. 5.** Relative pipeline displacement $U_p/U_{g,max}$ at time $t = 5$ s according to the semi-analytical, finite element modal and implicit dynamic analysis, considering an input ground displacement with frequency $f = 0.5$ Hz, and apparent wave propagation velocity $C_{ph} = 2000$ m/s, for various pipeline lengths: a) $L = 100$ m; b) $L = 1000$ m; c) $L = 10000$ m.

### 4.3 Frequency response analysis

To evaluate the system response as a function of the frequency of the ground motion, a frequency response analysis is performed using the developed semi-analytical model, considering different system lengths and soil stiffness. For comparison purposes, the duration of the harmonic ground motion is considered equal to 15 displacement cycles across the entire frequency range (0-20Hz), since the dynamic amplification depends on the number of applied loading cycles [8].
Figure 6a shows the variation of the dynamic amplification as a function of the ground motion frequency $f$, highlighting the resonance frequencies for the pipeline buried in compacted and poorly compacted backfill, for varying lengths $L$. The relative displacement $U_{p,max}/U_{g,max}$ is close to unity for low frequencies, increasing monotonically as the forcing frequency $f$ approaches the fundamental frequency of the system $f_1$, that is about 10.5 Hz and 3.3 Hz for the compacted and poorly compacted backfill condition, respectively. The dynamic amplification increases monotonically as a function of the forcing frequency and soil stiffness, from a minimum value of about 47 for poorly compacted backfill to 57 for the pipeline ($L$ = 100 m) buried in compacted soil (Figure 6b). This is because greater mode numbers $n$ are excited at higher resonant frequencies (Table 3), resulting in more modes contributing to the dynamic amplification. Although the resonant frequencies of the pipeline buried in compacted soil may fall outside the range of dominant frequencies of low-frequency earthquakes, they are relevant for other vibration sources, including traffic, giving a deeper understanding of the system dynamic behaviour.

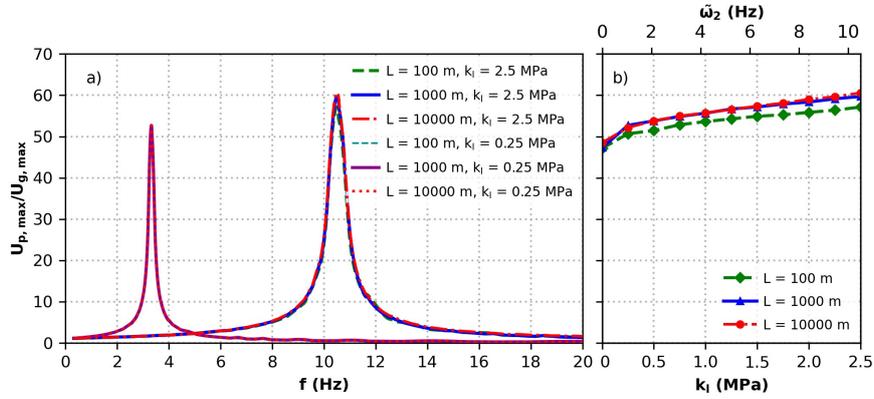

**Fig. 6.** Dynamic amplification of the soil-structure interaction for different system lengths $L$, as a function of the a) ground motion frequency $f$; b) soil stiffness $k_l$ ($\tilde{\omega}_2$).

**Table 3.** Maximum dynamic amplification $U_{p,max}/U_{g,max}$ and corresponding normalized frequency values $\omega(U_{p,max})/\tilde{\omega}_2$, considering various system lengths $L$ and soil stiffness $k_l$, characterized by the transition frequencies $\tilde{\omega}_i$, the total number of modes in the low frequency range $N(\omega_n \leq \tilde{\omega}_2)$, and the number $n$ of exited modal shape at resonance.

| $k_l$ (MPa) | $\tilde{\omega}_2$ (rad/s) | $\dfrac{\tilde{\omega}_1}{\tilde{\omega}_2}$ | $\dfrac{\tilde{\omega}_2}{\tilde{\omega}_2}$ | $\dfrac{\tilde{\omega}_3}{\tilde{\omega}_2}$ | $L$ (m) | $\dfrac{U_{p,max}}{U_{g,max}}$ | $\dfrac{\omega(U_{p,max})}{\tilde{\omega}_2}$ | $N(\omega_n \leq \tilde{\omega}_2)$ | $n$ |
|---|---|---|---|---|---|---|---|---|---|
| 2.5 | 66.08 | ≈1.0 | 1.0 | 90.2 | 100 | 57.15 | 0.9983 | 2 | 3 |
|  |  |  |  |  | 1000 | 59.71 | 0.9992 | 6 | 12 |
|  |  |  |  |  | 10000 | 60.60 | 0.9983 | 42 | 106 |
| 0.25 | 20.89 | ≈1.0 | 1.0 | 285.4 | 100 | 50.68 | 0.9993 | 2 | 1 |
|  |  |  |  |  | 1000 | 52.71 | 0.9993 | 3 | 5 |
|  |  |  |  |  | 10000 | 52.18 | 0.9993 | 14 | 35 |

## Conclusions

A new soil-structure interaction model is used to evaluate the dynamic response of Timoshenko beams on Winkler foundation subjected to transverse TGD. The validated semi-analytical model, accurately and efficiently evaluates the dynamic response of buried Timoshenko beams subjected to transverse ground vibration.

The closed-form analytical solution revealed that the vibration spectrum comprises four parts, separated by three transition frequencies. At each transition, the oscillatory characteristics of the modes change as a function of the system properties, leading to marked variations in dynamic amplification.

The pipelines buried in poorly compacted backfill exhibited a narrower and lower resonance bandwidth, because the lower modal density and the lower mode numbers excited at resonance, resulting in fewer modes contributing to dynamic amplification. Although the resonant frequencies of the pipeline buried in well compacted soil fall outside the range of dominant frequencies of low frequency earthquakes, they can be relevant for other vibration sources like high-frequency seismic, traffic and railway loadings.

The proposed method provides a robust analytical framework for evaluating the primary factors impacting the dynamic response of buried beams, giving a deeper understanding of the system response under various sources of ground vibration.